\documentclass[aps,prl,preprint,floatfix]{revtex4-1}
\usepackage{graphicx}

\begin{document}

\title{Enhanced transport at high plasma pressure and sub-threshold \\ kinetic ballooning modes in Wendelstein 7-X}

\author{P.~Mulholland$^1$}
\author{K.~Aleynikova$^2$} 
\author{B.~J.~Faber$^3$}
\author{M.~J.~Pueschel$^{1,4,5}$} 
\author{J.~H.~E.~Proll$^1$} 
\author{C.~C.~Hegna$^3$}
\author{P.~W.~Terry$^3$}
\author{C.~N{\"u}hrenberg$^2$} 

\affiliation{
$^1$Eindhoven University of Technology, 5600 MB Eindhoven, The Netherlands\\ 
$^2$Max-Planck-Institut f{\"u}r Plasmaphysik, 17491 Greifswald, Germany\\ 
$^3$University of Wisconsin-Madison, Madison, WI 53706, USA\\
$^4$Dutch Institute for Fundamental Energy Research, 5612 AJ Eindhoven, The Netherlands\\
$^5$Department of Physics \& Astronomy, Ruhr-Universit{\"a}t Bochum, D-44780 Bochum, Germany\\
}

\begin{abstract}
High-performance fusion plasmas, requiring high pressure $\beta$, are not well-understood in stellarator-type experiments. Here, the effect of $\beta$ on ion-temperature-gradient-driven (ITG) turbulence is studied in Wendelstein 7-X (W7-X), showing that subdominant kinetic ballooning modes (KBMs) are unstable well below the ideal MHD threshold and get strongly excited in the turbulence. By zonal-flow erosion, these sub-threshold KBMs (stKBMs) affect ITG saturation and enable higher heat fluxes. Controlling stKBMs will be essential to allow W7-X and future stellarators to achieve maximum performance. 
\end{abstract}

\maketitle

\newpage

\subsubsection{Introduction}

Turbulent transport is the leading cause of energy loss in modern magnetic confinement fusion devices. Electrostatic microinstabilities and turbulence have been studied in detail in both tokamaks \cite{Romanelli89, Cowley91, Beer95, Dimits00a, Connor06, Dannert05} and stellarators \cite{Xanthopoulos07, Zocco16, Zocco18a, Zocco20, Proll12, Faber15}. In high-performance regimes, the plasma will have substantial normalized plasma pressure $\beta \equiv \beta_\mathrm{e} = 8\pi n_\mathrm{e} T_\mathrm{e} / B_{\mathrm{ref}}^2$, where $n_\mathrm{e}$ is the electron density, $T_\mathrm{e}$ is the electron temperature and $B_\mathrm{ref}$ is the reference magnetic field. Note $\beta_\mathrm{total} = \beta_\mathrm{i} + \beta_\mathrm{e}$. This parameter is an indicator of reactor efficiency as reaction rates scale with $\beta^2$. However, finite $\beta$ can modify electrostatic modes and produce electromagnetic instabilities. Finite-$\beta$ studies have been carried out for tokamaks \cite{Connor78, Antonsen80, Tang80, Cheng82, Kotschenreuther86, Biglari91, Tsai93, Zonca96, Zonca99, Pueschel08a, Pueschel10a, Ishizawa13a, Ishizawa15a, Aleynikova17, Whelan18, Whelan19, Ishizawa19a, Ishizawa19b} and to a lesser extent for stellarators \cite{Ishizawa14a, Aleynikova18, Zocco18b, McKinney21, Aleynikova22}. Such studies aid in improving reactor efficiency, and are of increasing relevance as Wendelstein 7-X (W7-X) prepares to operate at high $\beta$. Therefore, a comprehensive understanding of how finite $\beta$ alters electrostatic turbulence and brings forth electromagnetic turbulence regimes, driven by instabilities such as the kinetic ballooning mode (KBM), is becoming increasingly important in fusion research. In this work, a gyrokinetic study is presented of finite-$\beta$ turbulence in W7-X geometry revealing a novel process involving enhanced transport due to KBM excitation well below the ideal MHD ballooning threshold $\beta_\mathrm{crit}^\mathrm{MHD}$. 

Depending on the device and regime, finite-$\beta$ studies have shown both positive and negative impacts on stability and turbulent transport. $\beta$ can suppress the growth of the ITG instability \cite{Rewoldt87, Kim93, Zonca99, Hirose00, Snyder01, Pueschel08a, Pueschel10a} and can reduce the transport levels associated with its turbulence \cite{Pueschel08a, Pueschel10a, Ishizawa14a, Ishizawa15a, Whelan18, Whelan19, McKinney21}. In other cases, high $\beta$ can have a disruptive influence on the efficacy of saturation, whereby zonal flows are eroded by radial motion of electrons in microstochastic fields \cite{Hatch12a, Hatch13, Terry13, Pueschel13b, Pueschel13c, Pueschel14a}. Electromagnetic stellarator turbulence studies have been carried out for NCSX \cite{Baumgaertel12}, LHD \cite{Ishizawa13a, Ishizawa14a, Ishizawa15a}, and more recently for HSX \cite{McKinney21}, in addition to linear studies of electromagnetic instability behavior for W7-X \cite{Aleynikova18, Aleynikova22} and global electromagnetic turbulence studies for W7-X \cite{Mishchenko21, Mishchenko22, Wilms21}. In LHD and HSX, it has been shown that a KBM-dominated turbulence regime may be more desirable than an ITG-dominated regime, based on the reduction in heat flux seen at large $\beta$. In contrast, this work reveals that a well-behaved KBM-dominated turbulence regime is not obtained in certain W7-X configurations, and instead, ITG-dominated turbulent fluxes increase while $\beta$ is far below the linear threshold of KBM dominance $\beta_\mathrm{crit}^\mathrm{KBM}$, and fail to saturate near $\beta_\mathrm{crit}^\mathrm{KBM}$. This enhancement is due to a newly-discovered process involving subdominant KBMs coupling to the zonal flow in ITG-dominated turbulence. The strong excitation of KBMs with increasing $\beta$ leads to a monotonic increase in transport. This is dissimilar to the previously reported non-zonal transition \cite{Pueschel13b, Pueschel13c, Pueschel14a}, which is marked by a sudden transport blow-up at a critical $\beta$ without KBM involvement. Thus, this work reports a physically distinct mechanism whose presence may be predicted from linear simulations alone. 

This result holds substantial implications for stellarator experiments attempting to achieve high-performance, given that high-$\beta$ operation is a key promise of the W7-X stellarator. Notably, W7-X's optimization relies on MHD stability \cite{Grieger90} and not on turbulence properties. Thus, this work motivates new approaches to stellarator optimization for electromagnetic instabilities and turbulence. 

\subsubsection{Simulation setup and dominant linear stability} 

Numerical studies are based on flux-tube simulations using the local $\textsc{Gene}$ code \cite{Jenko00}. Simulations use an MHD-optimized high-mirror configuration of W7-X \cite{Aleynikova18}. This configuration has mirror ratio $10\%$, rotational transform $\iota=1$ on axis and no horizontal shift \cite{Aleynikova22}. The neoclassical optimization of this configuration makes it a promising candidate for higher resilience to electromagnetic microinstabilities \cite{Aleynikova18}. Simulations use a flux tube centered at the outboard midplane of the bean-shaped plane \cite{Geiger14}, where one poloidal turn is sufficient for numerical convergence. The bean-shaped cross-section was chosen as it corresponds to the most MHD-unstable region of `bad' normal curvature \cite{Aleynikova18, Aleynikova22}. Furthermore, $T_\mathrm{i} / T_\mathrm{e}=1$, $n_\mathrm{i} / n_\mathrm{e} = 1$, $m_\mathrm{i} / m_\mathrm{e}=1836$, the radial position is $r/a = 0.7$ (normalized toroidal flux $s_0 = \psi(r)/\psi(a) = 0.5$), with normalized gradients $a/L_{T\mathrm{i}} = 3.5$, $a/L_{T\mathrm{e}} = 0$, $a/L_{n\mathrm{i}}=a/L_{n\mathrm{e}} = 1$. Here, $T_j$ is the temperature of species $j$, while $m_j$ is the mass, $r$ is the minor-radial coordinate, $a$ is the minor radius, and $L_{Tj} = -\left(d \ln T_j/ d r\right)^{-1}$ and $L_{nj} = -\left(d \ln n_j/ d r\right)^{-1}$ are the scale lengths of the temperature and density, respectively. At radial positions $s_0 = 0.3-0.6$, KBMs are expected to be present and detectable by experiment \cite{Aleynikova22}. Setting $a/L_{T\mathrm{e}}=0$ was chosen to maximise $a/L_{T\mathrm{i}}$ and decrease $\beta_\mathrm{crit}^{\mathrm{KBM}}$ below the MHD limit $\beta_\mathrm{crit}^{\mathrm{MHD}}$, given that the total sum of normalized gradients is held constant for a given $s_0$ and $\beta_\mathrm{crit}^{\mathrm{KBM}}$ depends more strongly on $a/L_{T\mathrm{i}}$ \cite{Cheng82, Pueschel08a, Aleynikova18}. However, we find that $\beta_\mathrm{crit}^{\mathrm{KBM}}$ in W7-X is largely dependent on the total sum of gradients, and not on a specific gradient \cite{Aleynikova17}. This is in line with the KBM being a pressure-gradient-driven instability, such that all gradients contribute to its growth. An additional study at $r/a = 0.76$ ($s_0 = 0.58$) including $a/L_{T\mathrm{e}} = 1.75$ is presented in Fig.~\ref{fig:ivs_evs_nl} (gray data, empty symbols). The inclusion of $a/L_{T\mathrm{e}}$ allows for stronger prediction of experimental performance. Notably, including $a/L_{T\mathrm{e}}$ yields analogous trends to $a/L_{T\mathrm{e}} = 0$. Thus, throughout this Letter and without loss of generality, focus is given to $a/L_{T\mathrm{e}} = 0$ (unless otherwise stated). Equilibria were created with the VMEC code \cite{Hirshman86}, using the pressure-profile procedure described in \cite{Aleynikova22}. For scans in $\beta$, distinct VMEC equilibria are used whose volume-averaged $\beta$ varies with the local $\beta$. Furthermore, a self-consistent sum of the normalized gradients is used, $\sum_{j=\mathrm{i,e}} \; a/L_{Tj} + a/L_{nj} = 5.5$ for radial position $s_0 = 0.5$ (and 7.25 for $s_0 = 0.58$). This $\alpha$-consistent approach ($\alpha$ being the normalized pressure gradient) ensures the equilibrium is self-consistent with the pressure gradient of the simulation. Deviations from this level of consistency can cause discrepancies in $\beta_\mathrm{crit}^\mathrm{KBM}$ \cite{Aleynikova18}.

\begin{figure}[h]
\centering
\includegraphics[width=8.6cm]{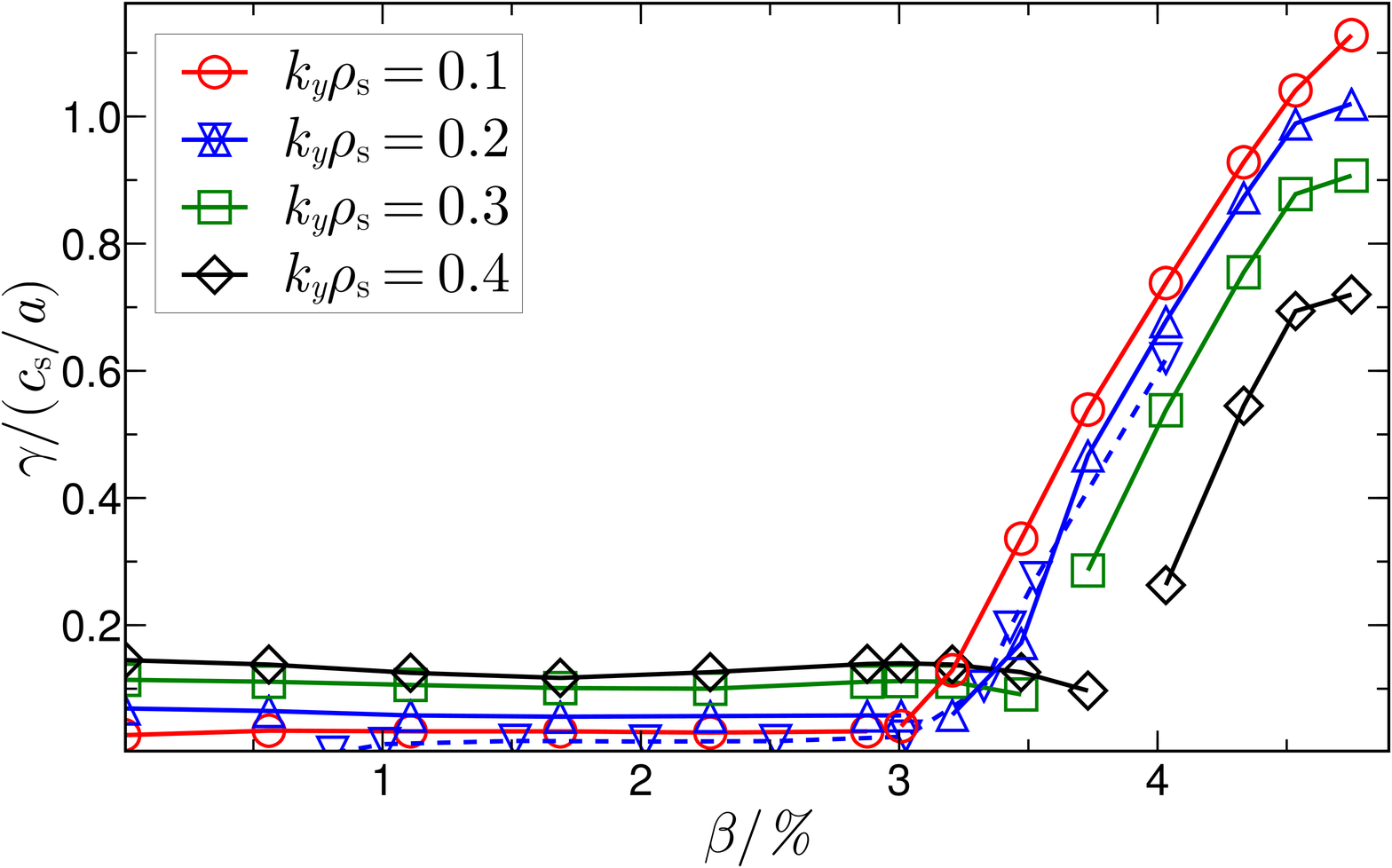}
\includegraphics[width=8.6cm]{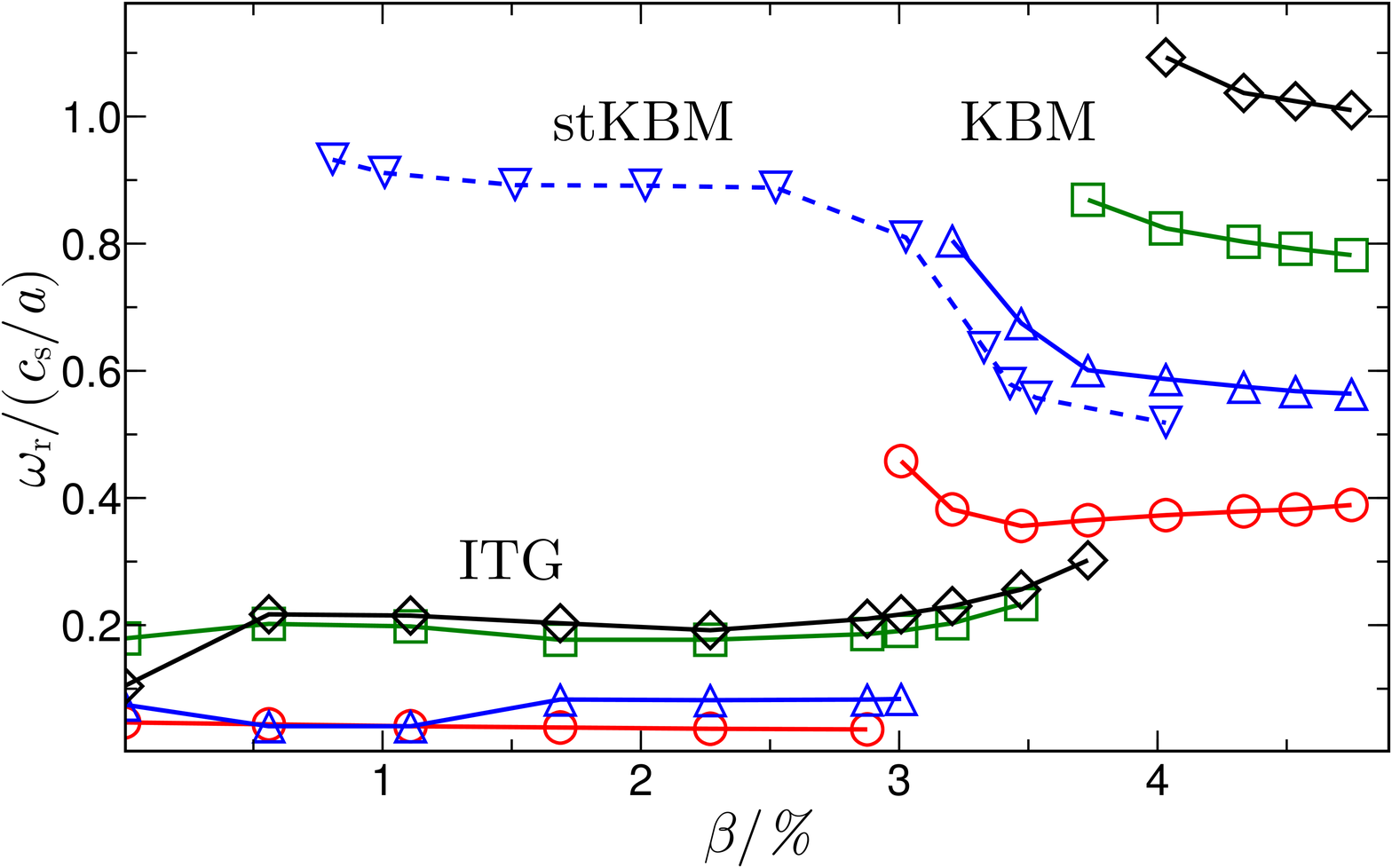}
\caption{Linear growth rates $\gamma$ and real frequencies $\omega_\mathrm{r}$ of the dominant mode as functions of $\beta$ for various $k_y$. The ITG instability dominates for $\beta \lesssim 3\%$ and is largely unaffected by $\beta$. Focusing on $k_y = 0.2$, the sub-threshold KBM (inverted blue triangles, dashed line) becomes unstable at $\beta_\mathrm{crit}^{\mathrm{KBM}} \approx 1\%$ and has a soft onset ($\gamma_{\mathrm{stKBM}}$ increases gradually with $\beta$) before becoming dominant $\left(\gamma_{\mathrm{stKBM}} > \gamma_{\mathrm{ITG}} \right)$ at $\beta_\mathrm{crit}^{\mathrm{KBM}} \approx 3\%$, where it transitions continuously into the fast-growing KBM (blue triangles, solid line).}
\label{fig:lin_ivs}
\end{figure}

Figure \ref{fig:lin_ivs} shows linear growth rates $\gamma$ and real frequencies $\omega_\mathrm{r}$ as functions of $\beta$, comparing different poloidal wavenumbers $k_y$ normalized to the inverse ion sound gyroradius $\rho_\mathrm{s}$. Discontinuities in frequency mark the regime transition between dominant ITG and dominant KBM, where the latter has a steeply increasing growth rate with $\beta$. These results are consistent with previous studies \cite{Pueschel08a, Pueschel10a, Ishizawa14a, Ishizawa15a, Aleynikova17, Aleynikova18, McKinney21}, with the exception of the absence of ITG suppression for $k_y \leq 0.4$ at $2\% < \beta < 3\%$, possibly a consequence of the self-consistent approach \cite{Ishizawa19a}. 

To obtain the approximate $\beta$ where $\gamma_{\mathrm{KBM}} > 0$, one commonly extrapolates $\gamma$ in the KBM-dominant regime to $\gamma=0$, which here yields a threshold $\beta \approx  3-3.5\%$. Also, given that the linear ITG is either suppressed or only moderately affected for $\beta < 3\%$, one may intuitively expect to see analogous behavior of the turbulence levels in this range of $\beta$. 

\subsubsection{Nonlinear simulations}

\begin{figure}[h]
\centering
\includegraphics[width=8.6cm]{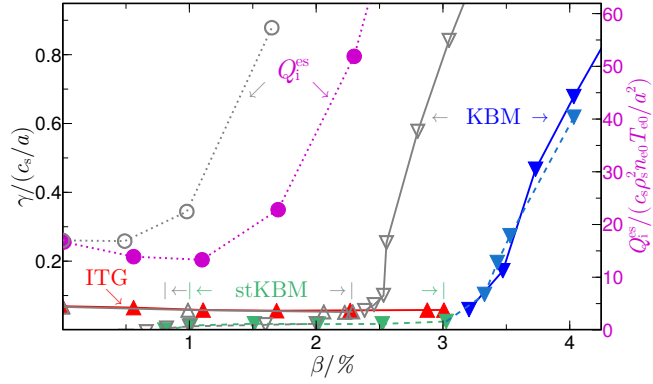}
\caption{Nonlinear ion electrostatic heat flux $Q_\mathrm{i}^\mathrm{es}$ (magenta circles, dotted lines) increasing for $\beta > \beta_\mathrm{crit}^{\mathrm{stKBM}}$ (stKBM: green inverted triangles, dashed lines) despite no similar scaling in the linear ITG growth rate (red triangles, solid lines). No saturation is achieved in the KBM-dominant regime (blue inverted triangles, solid and dashed lines). Gray data with empty symbols corresponds to $a/L_{T\mathrm{e}}=1.75$. Linear data corresponds to $k_y=0.2$.} 
\label{fig:ivs_evs_nl}
\end{figure}

Figure \ref{fig:ivs_evs_nl} shows the nonlinear electrostatic ion heat flux $Q_\mathrm{i}^\mathrm{es}$ averaged over the turbulent state for various $\beta$, together with linear growth rates for $k_y = 0.2$. Note that the electron and electromagnetic fluxes are small for $a/L_{T\mathrm{e}}=0$. For this W7-X configuration, $\beta_\mathrm{crit}^{\mathrm{KBM}} > 3\%$ for the fixed sum of gradients used here, while $\beta_\mathrm{crit}^{\mathrm{MHD}} \approx 3\%$. Therefore, one does not expect well-behaved saturated turbulence in the KBM-dominant regime. 

Convergence was achieved using $N_x = 256$ radial points, $N_{ky} = 24$ Fourier modes in $k_y$ with the smallest finite value 0.05, $N_z = 256$ gridpoints along the field line, $N_{v\parallel} = 32$ gridpoints for parallel velocity space, and $N_\mu = 12$ gridpoints for the magnetic moment. The heat flux is mostly unaffected for $0 < \beta < 1\%$, before sizeable increases manifest for $\beta > 1\%$. Fully-saturated turbulence is achieved up to $\beta \approx 2.6\%$, before fluxes increase rapidly without bound for $\beta > \beta_\mathrm{crit}^{\mathrm{MHD}} \approx 3\%$. The substantial increase in heat flux for $1\% < \beta < 3\%$ is unexpected and cannot be explained by the dominant-instability behavior alone.

\begin{figure}[h]
\centering
\includegraphics[width=8.6cm]{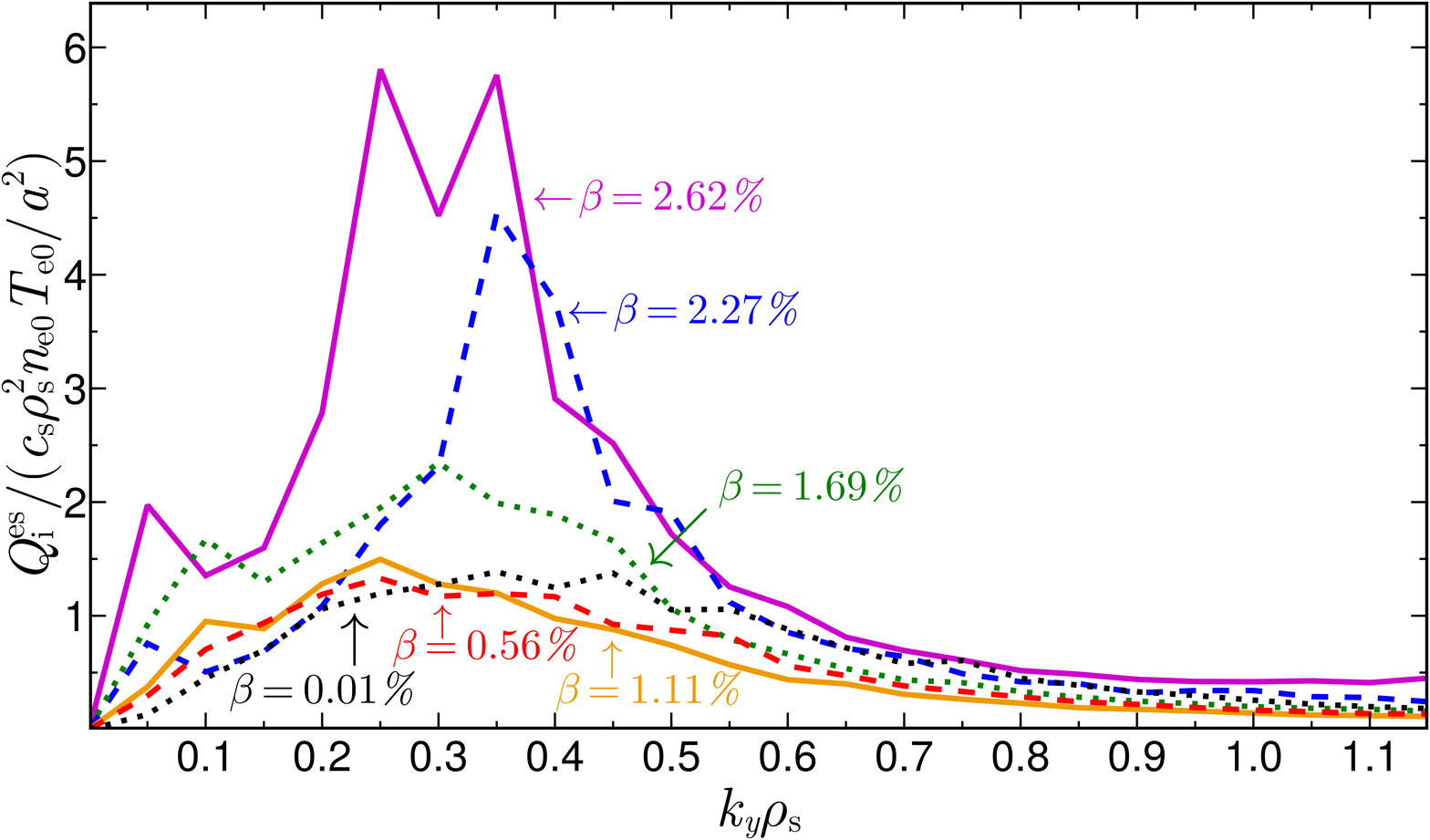}
\caption{Turbulent heat flux $Q_\mathrm{i}^\mathrm{es}(k_y)$. Spectra narrow at high $\beta$. The $\beta = 2.27\%$ and $2.62\%$ spectra have been rescaled by a factor of 1/2 for visibility. Spectra for $a/L_{T\mathrm{e}}=1.75$ (not shown) yield analogous trend.} 
\label{fig:nl_flux_spectra}
\end{figure}

The nonlinear heat-flux spectra, see Fig.~\ref{fig:nl_flux_spectra}, start broad for the electrostatic case $\beta = 0.01\%$ and become narrow at high $\beta$, peaking primarily at ITG-dominant $k_y = 0.3-0.4$, and to a lesser extent, at lower $k_y = 0.05-0.1$. Nonlinear frequencies (not shown), compared with the dominant linear frequencies are primarily in the ITG range at $\beta \approx 1\%$, but include a higher-frequency signature  -- characteristic of the KBM -- at $\beta > 2\%$. Nonlinear cross-phases (not shown) of the high-$\beta$ turbulence mostly resemble those found for dominant linear ITGs at high $\beta$. These diagnostics suggest that the high-$\beta$ turbulence is driven by a mixture of ITGs and KBMs. 

To understand what processes could be causing the increase in heat flux, consider the saturation mechanism at low $\beta$, where zonal flows play a critical role (see Refs.~\cite{Diamond05, Itoh06} and references therein). Zonal flows are excited to substantial amplitudes for $\beta \lesssim 1\%$. However, at $\beta \approx 2\%$, substantial zonal-flow reduction occurs and streamer-like structures form that span $\approx$ 100$\rho_\mathrm{s}$, accompanied by a substantial drop in the normalized zonal potential $|\Phi_\mathrm{zonal}|^2/ |\Phi_\mathrm{nonzonal}|^2$. Here, $\Phi_\mathrm{zonal} \equiv \Phi(k_y=0)$ and $\Phi_\mathrm{nonzonal} \equiv \sum_{k_y > 0} \Phi(k_y)$. At $\beta = 1.11\%$, $|\Phi_\mathrm{zonal}|^2 / |\Phi_\mathrm{nonzonal}|^2 \approx 9$, while at $\beta = 2.27\%$, $|\Phi_\mathrm{zonal}|^2 / |\Phi_\mathrm{nonzonal}|^2 \approx 1$. This supports the notion that the increasing heat flux with $\beta$ is due to a reduced zonal flow.  

To address whether zonal-flow generation has been interrupted or the zonal flows have been eroded, the effect of $\beta$ on secondary instability is measured \cite{Rogers00, Pueschel13a}. Zonal modes ($k_x \neq 0$ and $k_y = 0$) grow exponentially via a three-wave interaction involving the ITG streamer ($k_x = 0$ and $k_y \neq 0$) -- held constant in time -- coupled with sidebands ($k_x \neq 0$ and $k_y \neq 0$). These simulations consider 17 complex modes in $k_x$ centered around $k_x = 0$ and at $k_y = 0.4$ and 0.8, which correspond to strong linear growth. Secondary-instability analysis constitutes a simplified model involving a subset of the mode couplings present in the turbulence, but mirrors the zonal-flow-related energetics of the turbulent system \cite{Pueschel13c}. Zonal-flow growth rates are measured at each $\beta$ separately, where the amplitudes are normalized, following the standard approach \cite{Pueschel13a}. Increasing $\beta$ from $1.11\%$ to $2.27\%$ slightly reduces zonal-mode growth rates by $\mathcal{O}(10\%)$, which is unlikely to explain the significant reduction in zonal-flow amplitude. 

Secondly, magnetic stochasticity is quantified at low and high $\beta$, which can reveal if radial motion of electrons is able to erode the zonal flow \cite{Hatch12a, Hatch13, Terry13, Pueschel13b, Pueschel13c, Pueschel14a}. This is done by evaluating the field-line diffusivity \cite{Pueschel13c} 

\begin{equation}
D_{\mathrm{fl}}(l,p) = \frac{[\Delta r(l,p)]^2}{2 \pi q_0 a (p+1)},
\end{equation}

which measures the radial displacement $\Delta r(l,p) = r(l,p) - r(l,0)$ of field-line $l$ after poloidal turn $p$, where $q_0$ is the safety factor. Here, diffusivity is averaged over the quasi-stationary state using 10 poloidal turns, showing an increase from 0.0053 $\rho^2_{\mathrm{s}} / a$ to 0.11 $\rho^2_{\mathrm{s}} / a$ as $\beta$ increases from 1.11\% to 2.27\%. Furthermore, at $\beta = 2.27\%$, the zonal-flow decay time (1.23 $a / c_\mathrm{s}$) has become shorter than the turbulent correlation time (3.92 $a / c_\mathrm{s}$), supporting the notion that zonal-flow erosion is a dynamically relevant effect \cite{Pueschel13b, Terry13}. Therefore, the zonal flow is much harder to maintain with increasing $\beta$, lowering its ability to saturate turbulence.

Typically, stochasticity is produced by modes with even parity in the magnetic vector potential $A_\parallel$ \cite{Nevins11, Hatch12a, Hatch13}. Tearing parity (odd parity in $\Phi$, even parity in $A_\parallel$) is seen in the present case when averaging over the quasi-stationary state. In particular, a proper orthogonal decomposition reveals that the tearing-parity contribution to $A_\parallel$ increases at $k_y = 0.1-0.25$ when $\beta > 2\%$. 

\subsubsection{Eigenmode spectra and projection} 

At different $k_y$, eigenvalue calculations are performed to study the subdominant spectrum. Resolutions for the eigenvalue calculations are the same as for nonlinear simulations, with the exception of $N_x = 29$. To assess the excitation of various eigenmodes in the turbulence, the non-adiabatic part of the perturbed distribution function $g_{\mathrm{NL}}$ is projected onto eigenvector $g_j$ at a given time, thus obtaining the projection \cite{Hatch13, Pueschel16, Faber18}

\begin{equation}
p_j=\frac{\left|\int \mathrm{d}\theta g_j^* g_{\mathrm{NL}}\right|}{\left(\int \mathrm{d}\theta \left|g_j\right|^2 \int \mathrm{d}\theta \left|g_{\mathrm{NL}}\right|^2\right)^{1 / 2}},
\label{eq:proj}
\end{equation}
\vspace{0.1cm}

giving the excitation of the $j$-th eigenmode, which is then averaged over the turbulent state. Here, $\theta$ is the extended ballooning angle \cite{Connor78, Candy04}, and a summation over both particle species is implicit. If $p_j = 1$, the nonlinear state is captured by a single eigenmode, whereas a mode with $p_j = 0$ is orthogonal to the nonlinear state. The set of eigenmodes itself is not orthogonal, and generally $\sum_j p_j > 1$. 

\begin{figure}[h]
\centering
\includegraphics[width=8.6cm]{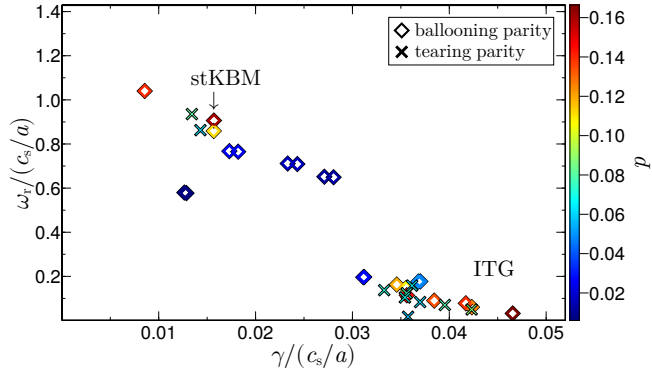}
\caption{Linear eigenmode spectrum at $k_y = 0.2$ and $\beta = 1.11\%$, with nonlinear excitation $p$ shown in color. The ITG cluster resides at $\omega_\mathrm{r} < 0.3 \ c_\mathrm{s}/a $ and the KBM cluster resides at $\omega_\mathrm{r} > 0.3 \ c_\mathrm{s}/a$. The sub-threshold KBM is strongly excited, but less than the highest-projection ITG.} 
\label{fig:proj_ky2b11}
\end{figure}

\begin{figure}[h]
\centering
\includegraphics[width=8.6cm]{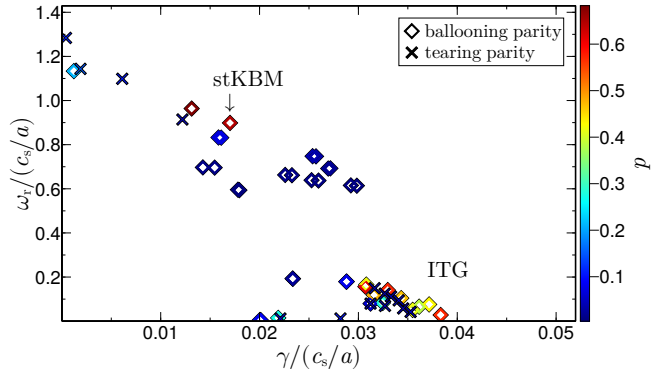}
\caption{Linear eigenmode spectrum at $k_y = 0.2$ and $\beta = 2.27\%$ with nonlinear excitation $p$ shown in color. The ITG cluster resides at $\omega_\mathrm{r} < 0.3 \ c_\mathrm{s}/a$ and the KBM cluster resides at $\omega_\mathrm{r} > 0.3 \ c_\mathrm{s}/a$. The sub-threshold KBM is more strongly excited than the highest-projection ITG.} 
\label{fig:proj_ky2b22}
\end{figure}

Figures \ref{fig:proj_ky2b11} and \ref{fig:proj_ky2b22} show the eigenspectra and projections at $k_y = 0.2$ for $\beta = 1.11\%$ and $2.27\%$, respectively. Two clusters of modes can be identified: an ITG cluster ($\omega_\mathrm{r} < 0.3 \ c_\mathrm{s}/a$) and a KBM cluster ($\omega_\mathrm{r} > 0.3 \ c_\mathrm{s}/a$). The projections reveal several large-amplitude subdominant modes ($\gamma < \gamma_{\mathrm{max}}$) within both clusters. KBM excitation increases as $\beta$ increases from $1.11\%$ to $2.27\%$, where in the latter case, some modes in the KBM cluster surpass the highest-projection ITGs. The relative orthogonality between all modes is measured qualitatively using a modified form of Eq. \ref{eq:proj} (replacing $g_{\mathrm{NL}}$ with $g_k$, $k \neq j$). The mode of interest in the KBM cluster (stKBM in Fig.~4 and 5) has $< 17\%$ similarity to the high-projection ITGs. In contrast, another high-projection mode in the KBM cluster has $\approx 58\%$ similarity and $\approx 30-38\%$ similarity to the high-projection ITGs and stKBM, respectively. Thus, the large excitation of this mode is unlikely to be primarily associated with its direct influence on the turbulence, whereas the highly-excited and largely-distinct stKBM and ITGs are expected to play a unique and significant role. 

A more fine-grained study has been carried out for incremental increases in $\beta$, revealing that a single KBM becomes unstable for $\beta \approx 1\%$ (the sub-threshold KBM, destabilized at $\beta_\mathrm{crit}^{\mathrm{stKBM}}$) and has a slowly increasing $\gamma$ up to $\beta_\mathrm{crit}^{\mathrm{KBM}} \approx 3\%$, at which point $\gamma$ increases rapidly. This evolution of the KBM from soft onset to rapid growth is visible in Fig.~\ref{fig:lin_ivs} (blue inverted triangles), where for $\beta > 3\%$ this single KBM has split off from the KBM cluster and has become the dominant instability. This novel property of the KBM having a soft onset -- whilst being strongly excited in the turbulence -- implies that one cannot reliably use extrapolation in Fig.~\ref{fig:lin_ivs} to obtain the threshold where the KBM first impacts the system dynamics. 

Eigenspectra at $k_y = 0.3$ and $0.4$ (not shown) contain analogous ITG and KBM clusters; however, the excitation of KBMs remains lower than of ITGs for both $\beta = 1.11\%$ and $2.27\%$ at these wavenumbers. Thus, the primary impact of KBMs on the turbulence is to act as an unprecedented catalyst that enables ITGs to produce larger fluctuations and fluxes at $k_y = 0.3-0.4$. A likely candidate is the nonlinear energy transfer between zonal flows and stable tearing-parity modes with $0 < k_y < 0.2$ aided by sub-threshold KBMs, which causes the observed magnetic stochasticity, zonal-flow suppression, and flux increase. All unstable tearing-parity eigenmodes are found to have very low excitation ($p < 0.08$), such that they are unlikely to have significant impact on the turbulence. 

As mentioned earlier, $\beta_\mathrm{crit}^\mathrm{KBM}$ is largely unaffected by the relative balance of normalized gradients, but is set by their total sum. Fig.~\ref{fig:ivs_evs_nl} shows the reduction of $\beta_\mathrm{crit}^\mathrm{KBM}$ when including $a/L_{T\mathrm{e}}=1.75$. Notably, $\beta_\mathrm{crit}^\mathrm{stKBM}$ is also reduced in this case, showing the similar response of stKBM and dominant KBM to an increased sum of normalized gradients. 

The low threshold $\beta_\mathrm{crit}^{\mathrm{stKBM}} < \beta_\mathrm{crit}^{\mathrm{MHD}}$ may be attributable to the low magnetic shear of W7-X, $\hat{s} = -(r/\iota) d\iota / dr \ll 1$ (relatedly, $\beta_\mathrm{crit}^{\mathrm{KBM}} \propto |\hat{s}|$), in combination with an ion-magnetic-drift resonance whereby thermal ions exchange energy with the drift wave \cite{Hirose95, Hirose96, Aleynikova17, Chen18, McKinney21}. An in-depth study on the effects of magnetic shear on KBMs in W7-X will be presented in future work.
 
\subsubsection{Conclusions}

This Letter presents a novel process whereby electromagnetic turbulence is shown to have a significant and detrimental impact on energy confinement in W7-X for a high-performance scenario. In particular, the excitation of subdominant KBMs is shown to catalyze a form of zonal-flow degradation, allowing for enhanced ITG-driven transport as $\beta$ is gradually increased. 

This work examines the effect of high $\beta$ on the ITG instability and defines distinct thresholds for KBM destabilization ($\beta_\mathrm{crit}^\mathrm{stKBM} \approx 1\%$) and the much higher threshold of KBM dominance ($\beta_\mathrm{crit}^{\mathrm{KBM}} \approx 3\%$). Nonlinear simulations show increasing heat fluxes with $\beta$, starting from $\beta \approx 1\%$ and continuing up to the MHD limit $\beta_\mathrm{crit}^{\mathrm{MHD}} \approx 3\%$, despite insensitivity of linear ITG growth rates to $\beta$ and heat fluxes peaking at ITG-dominant wavenumbers. This is caused by zonal-flow reduction stemming from the eroding effect of magnetic stochasticity, which increases substantially with $\beta$. Projection analyses reveal the strong excitation of stKBMs in the turbulence as $\beta$ increases. Therefore, nonlinear energy transfer -- catalyzed by stKBMs -- between the zonal flow and stable tearing-parity modes is the likely cause of the observed increase in stochasticity, zonal-flow reduction, and heat-flux increase. Further investigation into this nonlinear process will be reported in future. 

Including $a/L_{T\mathrm{e}}=1.75$ yields analogous results to $a/L_{T\mathrm{e}}=0$, with the modification that both stKBM destabilization and heat-flux increase occur at a reduced $\beta \approx 0.7\%$. This indicates that KBM-enhanced transport may be even more prevalent in experimentally-relevant scenarios. 

The results reported here for W7-X may prove to be broadly applicable to low-magnetic-shear stellarators, and have significant implications regarding their maximum achievable $\beta$ and thus reactor performance. Moreover, this calls into question the viability of seeking configurations with high ideal MHD ballooning limits in stellarator optimization efforts. Optimizing magnetic geometries for high fusion performance will likely require accounting for stKBM activity via linear eigenvalue simulations. Controlling stKBMs will thus be essential for stellarators, and will pave the way for successful high-performance discharges in future reactor designs. 

\newpage

\begin{acknowledgments}

\subsubsection{Acknowledgments}

This work has been carried out within the framework of the EUROfusion Consortium, funded by the European Union via the Euratom Research and Training Programme (Grant Agreement No 101052200 — EUROfusion). Views and opinions expressed are however those of the author(s) only and do not necessarily reflect those of the European Union or the European Commission. Neither the European Union nor the European Commission can be held responsible for them.
\end{acknowledgments}

\newpage

\bibliographystyle{apsrev4-1}
\bibliography{refs}

\end{document}